*-20221130*

# Statistics for Spatially Stratified Heterogeneous Data

Jin-Feng Wang[1,2*], Robert Haining[3], Tonglin Zhang[4], Cheng-Dong Xu[1], Mao-Gui Hu[1]

[1]LREIS, Institute of Geographic Sciences and Natural Resources Research, Chinese Academy of Sciences, Beijing 100101, PR China

[2]University of Chinese Academy of Sciences, Beijing, PR China

[3]Department of Geography, University of Cambridge, Cambridge CB2 3EN, UK

[4]Department of Statistics, Purdue University, IN 47907-2066, USA

*Corresponding author

JFW: wangjf@Lreis.ac.cn

RPH: rph26@cam.ac.uk

TLZ: tlzhang@purdue.edu

CDX: xucd@Lreis.ac.cn

MGH: humg@Lreis.ac.cn

Word count in abstract: 193

Word count in text: 8293

Number of Tables: 3

Number of Figures: 10

# Statistics for Spatially Stratified Heterogeneous Data

**Abstract**. Spatial statistics is dominated by spatial autocorrelation (SAC) based Kriging and BHM, and spatial local heterogeneity based hotspots and geographical regression methods, appraised as the first and second laws of Geography (Tobler 1970; Goodchild 2004), respectively. Spatial stratified heterogeneity (SSH), the phenomena of a partition that within strata is more similar than between strata, examples are climate zones and landuse classes and remote sensing classification, is prevalent in geography and understood since ancient Greek, is surprisingly neglected in Spatial Statistics, probably due to the existence of hundreds of classification algorithms. In this article, we go beyond the classifications and disclose that SSH is the sources of sample bias, statistic bias, modelling confounding and misleading CI, and recommend robust solutions to overcome the negativity. In the meantime, we elaborate four benefits from SSH: creating identical PDF or equivalent to random sampling in stratum; the spatial pattern in strata, the borders between strata as a specific information for nonlinear causation; and general interaction by overlaying two spatial patterns. We developed the equation of SSH and discuss its' context. The comprehensive investigation formulates the statistics for SSH, presenting a new principle and toolbox in spatial statistics.

## 1. Introduction

Statistical theory was initially developed for independent and identically distributed (iid) observations from a population (often obtained as a result of carrying out a series of experiments) with the aim of testing hypotheses and estimating model parameters. However spatial data, data where each observed value has a geo-reference, are not independent. Observations on a variable of interest that are close to each other in geographical space tend to be similar, and such data are referred to as spatially autocorrelated (SAC) in the literature. If we can assume that the observations are collected from a distribution where underlying spatial variation can be described by a statistical model with a small number of parameters, then statistical inference becomes possible (see for example Haining 2003; Christakos 1992). Modelling spatial dependence, through some form of spatial autoregressive model for small area ("lattice") data, or permissible semi-variogram for geostatistical data, lies at the heart of many branches of spatial statistics (Cliff and Ord 1981; Cressie 1993). The property of spatial dependence in small area data is not independent of the size of the areal units (see for example Kolasa and Rollo 1991). For data from a continuous surface, the shape and properties of the empirical semi-variogram often depend on the sampling interval, that is, the distance between sample points (Issaks and Srivastava 1989).

The ability to model spatial dependence opens a door to undertaking many forms of spatial analysis, such as, spatial interpolation where sample data are used to construct a map of surface variation and its associated error by interpolating data values at locations where no data were collected (Matheron 1963; Griffith 2003). Spatial interpolation at a location is carried out by weighting data values from nearby geographical locations where the weights, in the case of kriging, are estimated from the semi-variogram. The underlying principal can be described as one of "borrowing nearby *data* values" for the purpose of estimating other *data* values in order to construct a map. Models of spatial variation can also be used to improve the precision of small area parameter estimates by a similar process of "borrowing information", such as, using data from neighbouring areas to estimate the *parameter* of any given area. The underlying assumption that allows this form of information borrowing is that neighbouring *parameter* values are similar so that the information contained in the data from neighbouring areas can be "borrowed" to estimate the unknown parameter (Rao 2003; Haining and Li 2020). The exploratory and explanatory analysis of spatial data recognizing their non-independence property underwent significant advances from the 1950s onwards and the reader interested in judging these early developments is referred to Ripley (1981). Rao (2014) reviews three approaches to small area estimation, including multilevel modeling (Goldstein 1996; Fazio and Piacentino 2010; Aiello and Ricotta 2016); and Fotheringham and Rogerson (2009) and Lloyd (2010) provide more recent, geographically oriented overviews.

This paper is concerned with the development of methods that can address another frequently encountered property of spatial data: spatial heterogeneity. Spatial heterogeneity is a term used in statistics to indicate that one or more statistical characteristics of interest are not the same across all subsets of the population (Everitt and Skrondal 2010, p204). The presence of spatial heterogeneity violates the second part of the iid assumption - observations are not "identically distributed" (Osborne et al. 2007; Dormann et al. 2007). Particularly, if our study region is large and physically or socio-economically diverse or our study region is observed in high spatial resolution, then the assumption that all subsets of our data have the same statistical characteristics is likely to be invalid. The assumption that *not* all subsets of our data have the same statistical characteristics, may be a safer starting point.    Dutilleul (2011) describes three types of often encountered spatial heterogeneity: heterogeneity in the mean, heterogeneity in the variance (heteroscedasticity), and heterogeneity associated with the autocorrelation structure in the data (Dutilleul 2011, p.20-21).

Heterogeneity in the mean (or *first order* heterogeneity) can be checked by a chi-square test in the case of count data or an ANOVA test in the case of continuous valued data (see Haining and Li 2020, chapter 6).    Heterogeneity in the mean may arise as a result of variation in the levels of a set of independent variables.   Providing those variables can be correctly specified and the relationship is structurally stable (model parameters are constant over the study area) a regression model can be fitted to account for heterogeneity in the mean.   However, if regression parameters vary spatially, then the relationship is said to be "structurally unstable" or spatially varying. Geographically weighted regression (Fotheringham et al. 2000), spatially varying coefficients and profile regression models will allow the data analyst to explore and model this form of heterogeneity in the mean (Lloyd 2010; Haining and Li, 2020, chapters 6 and 9).

Heterogeneity in the spatial autocorrelation structure of the data (or, together with heteroscedasticity, what can be termed *second order* heterogeneity) is also commonly encountered. This form of heterogeneity often takes the form of *localized clusters* of high (or low) values – in contrast to the *global* property of high (or low) values having a general propensity to be found together which is referred to as spatial autocorrelation.    A number of statistical tests are available to detect this form of heterogeneity including the class of local indicators of spatial autocorrelation or LISAs (Anselin 1995), $G_i$ and $G_i^*$ statistics (Getis and Ord 1992) and spatial scan statistics (Kulldorff 1997).    They are widely used although some of these tests are subject to the multiple testing problem.    Simply expressed, the more simultaneous tests that are run on a sample (which occurs, for example, when carrying out *n* tests, one for each of the *n* areas partitioning a region) the greater the probability of rejecting the null hypothesis for at least one of these *n* tests.    The probability of making a Type I error (rejecting the null hypothesis when it is true) exceeds the decision rule (e.g. 5% or 10%) chosen by the researcher.    For further details, see Haining and Li 2020, chapter 6.

Another important form of heterogeneity is spatially stratified heterogeneity (SSHy). It arises where an area comprising sets of contiguous spatial units can be partitioned into distinct spatial segments (strata) where *within* each stratum (each comprising a number of spatial units) the mean value of a variable or the association between variables is the same so that each stratum displays within-stratum homogeneity. These statistical characteristics may not be the same when compared with other strata so that collectively the strata display between-strata heterogeneity (Wang et al. 2016). This problem seems to have received less systematic attention than the other forms of local spatial heterogeneity described above. Part of the reason for this may lie in meeting the challenge of identifying homogeneous zones – a topic to which we will return in section 3 of this paper. SSHy is a source of confounding when a global model assuming homogeneity is applied to a SSH population (Simpson Paradox). Even if heterogeneity is recognized, there may be insufficient data to provide good estimates of the parameters in each stratum using traditional estimation methods – referred to as the data sparsity problem, or even biased sample problem when not al strata are sampled (Wang et al. 2018; Xu et al. 2018; Haining and Li 2020).

This article aims through theory and with reference to previous studies to promote the framework of statistics for SSHy. The paper is structured as follows. In section 2, we consider the principle statistical challenges that arise when working with SSH populations. Section 3 introduces the *q*-statistic for testing SSH populations which is where we address the challenge of identifying spatially stratified homogeneous zones. In section 4, we give examples of statistical inference applied to SSH populations. In section 5 we draw the reader's attention to a number of applications where SSHy is an issue for data analysis because of the nature of the scientific problem. We comment here on the relationship between SSHy and the methods of this paper on the one hand and the modifiable areal units problem on the other. We draw some final conclusions and directions for future work in section 6.

## 2. Statistical Challenges When Working with Spatially Stratified Heterogeneous (SSH) Populations

In this section we present examples of statistical problems that can arise if SSHy goes undetected in the course of a statistical analysis.

*2.1 Unrepresentative samples and poor quality estimates*

By an *unrepresentative sample* we mean that the histogram of the sample differs significantly from that of the population from which it is drawn. If SSHy is present but not recognized when designing a sample this can have a number of consequences. For example, the sample size may be too small to provide adequate coverage of all the strata. There may be no samples in some strata (so no estimate of the parameter of interest is possible) or so few samples that either no estimate of error can be computed (sample of size 1) or the estimate has an unrealistically small or large error variance (samples of size 2 or 3 for example). If a large national survey is carried out but stratification is not considered, then the foregoing problems can be addressed using hierarchical modelling which involves the "borrowing information" methodology referred to in section 1 (see Haining and Li 2020, chapters 7 and 8).[1]

Bradley et al (2021) reported that US vaccine uptake is overestimated by sampling Facebook users only. Behind which is that the US citizens' attitudes to the vaccine are related to their attitudes to the Facebook, i.e. the population is (spatial) stratified heterogeneity. Figure 1 shows one more biased sampling problem: the distribution of climatic zones within China and the distribution of meteorological stations in China in 1900. The inadequacy of that early network of stations for providing statistics for the whole country is all too evident. If the goal of inference is to obtain an estimate of the mean of some attribute (for example, average rainfall for the whole of China), then a better option is to introduce spatial stratification into the sample where the strata are chosen to reflect China's climatic zones. Supplementary information, if available, may be called upon to reduce the effects of any intra-stratum sample bias (Heckman 1979; Wang et al. 2018). For example, the number and spacing of samples in any one zone will depend on intra-zone attribute variation and the strength of spatial autocorrelation since if spatial autocorrelation is present then there is little purpose served in choosing sample sites close together. Rodriguez-Iturbe and Mejia (1974) discuss the challenges associated with setting up a new rainfall monitoring network in geographical space whilst O'Connell et al. (1979) study the problem of rationalizing an existing rain

---

[1] It should be remarked that even when a spatially distributed population is homogeneous (i.e. is not SSH) stratification of the geographical area is often recommended in order to obtain an efficient parameter estimator (e.g. of the population mean). However this is because of the presence of spatial dependence not in order to counter the effects of the possibility of SSHy. See Ripley 1981, p.19-27; Dunn and Harrison 1993.

gauge network by deleting sites where there is data redundancy.

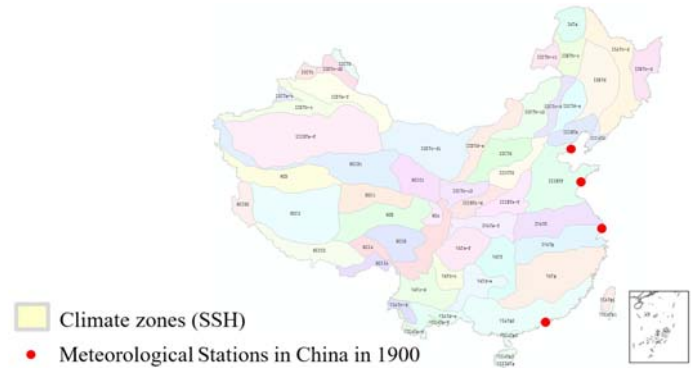

**Figure 1**. Climate zones in China and meteorological stations in 1900

*2.2 Confounded relationships*

The presence of SSHy undermines inferences that are based on global statistics that disregard any stratified heterogeneities in the population (Lindley and Novick 1981; Hox 2010, p3). For example, Xu et al (2011) using bubonic plague records and climate records for the period 1850 to 1964 show that the intensity of plague in China shows no clear evidence of an association with wetness levels until the data are broken down into the Northern and Southern halves of the country. Figure 2a shows a plot based on data for the whole country whilst Figures 2b and 2c partition the data into North China and South China. Xu et al. (2011) conclude that plague carrying rodent communities respond differently to higher levels of precipitation in arid Northern China compared with humid Southern China.

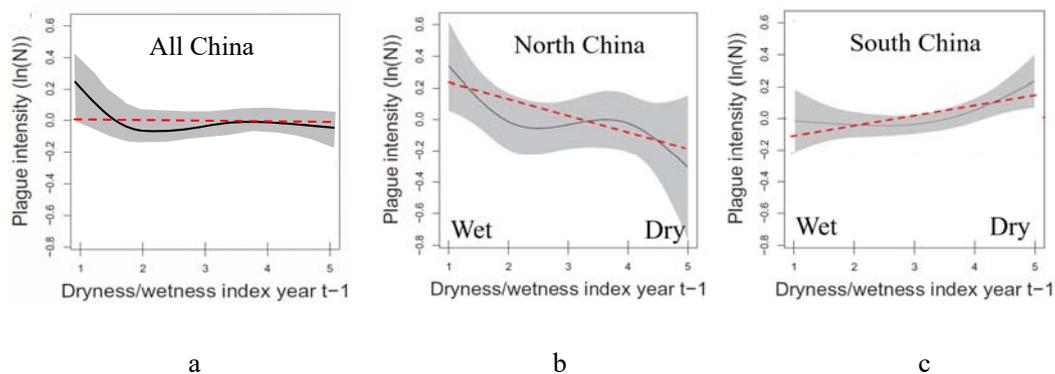

a                  b                  c

**Figure 2**. The relationship between plague intensity and wetness levels. (a) all China (b) only North China; (c) only South China. ((b) and (c) are reproduced with the permission of Xu et al. 2011). The solid curves are the GAM (Generalized Additive Model)' associations, their linear

trends are represented by dotted red lines; plague intensity refers to the number of plague cases (N) per year.

Similarly, if the population is SSHy, any global model based on a pooling of datasets would give rise to errors in prediction for both entire area and in at least one or more subsets without the user necessarily knowing which subsets. In a special issue of the *Canadian Water Resources Journal*, overview processes that generate flooding in Canada, Buttle et al. (2016) reported that a global model cannot provide accurate flood predictions for both entire and different regions of Canada, subjecting to different climatic regimes so different flood generating processes including flooding generated by snowmelt, rain-on-snow and rainfall as well as groundwater flood processes and flooding induced by storm surges, ice jams and urban flooding.

**3. Equations of Spatially Stratified Heterogeneity**

In this section we illustrate SSH in the real world (Figure 3) and its test in statistics (Figure 4), which is then decomposed in equations. This facilitates the investigation of its function ($L$, $P$), the number of strata ($L$) and the way to partition of the strata ($P$).

Figure 3 illustrates a range of maps displaying data values with different degrees of spatial structure. Figure 3a represents a map that shows no evidence of stratification. Figure 3b illustrates a map with some small degree of spatial structure (some spatial autocorrelation between neighbouring values) but still showing no evidence of any stratification. Figure 3d shows a map with two sharply defined strata and with strong homogeneity within each of the strata (little or no intra-strata variability). Figure 3c depicts an "imperfectly stratified heterogeneous" map where data values are well structured in space but there is some within-strata variation and the boundary lines (in red) between the three strata, are somewhat fuzzy – there appear to be three spatial regimes but the position of the boundary lines could be debated. In practice 3c represents the most likely scenario. Attached to each map in Figure 3 is a notional value of what is termed the *q*-statistic, which provides a measure of SSHy. We now discuss this statistic.

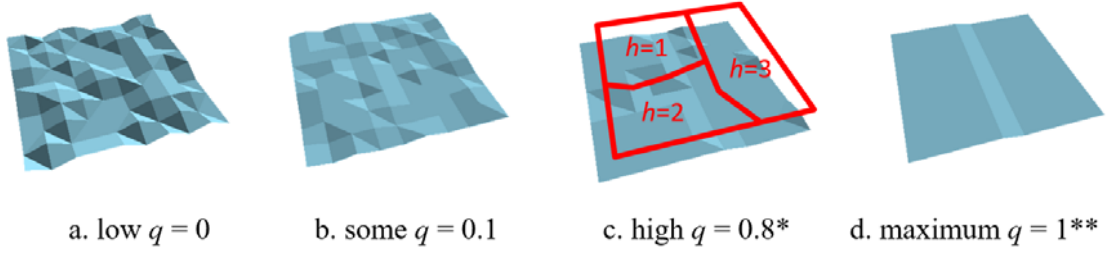

a. low $q = 0$     b. some $q = 0.1$     c. high $q = 0.8$*     d. maximum $q = 1$**

**Figure 3**. Maps displaying different amounts of spatially stratified heterogeneity (see text) together with the corresponding values of the $q$-statistic. * and ** stand for 0.05 and 0.01 levels of significance, respectively.

*3.1 A Model for a SSH population*

A SSH map is composed of two levels: a unit level $\{i\}$ and a higher strata level $\{h\}$, and is modelled by a pair of equations as shown below (Thompson 2012, p149; Goldstein 2011, p19):

$$\text{Individual equation: } Y_{hi} = M_h + \varepsilon_{hi} \text{ with } \varepsilon_{hi} \sim^{iid} N(0, \sigma_h^2), h = 1, …, L \quad (1)$$

$$\text{Strata equation: } M_h = M + u_h \text{ with } u_h \sim N(0, \sigma_B^2) \quad (2)$$

where $Y_{hi}$ denotes a random variable at site $i$ in stratum $h$, $M_h = E(Y_{hi} | M_h)$ for $\forall i$. $M = \Sigma_h^L W_h M_h$, $W_h = N_h/N$, where $N_h$ and $N$ are the numbers of units in stratum $h$ and the population respectively. $\sigma_h^2 = E(Y_{hi} - M_h|M_h)^2$, $\sigma_B^2 = V[E(Y_{hi}|M_h)] = E(M_h - M)^2$. To illustrate $Y_{hi}$ might refer to annual temperature at a meteorological station $i$ in climate zone $h$, $M_h$ might refer to annual temperature averaged over climate zone $h$, $M$ refers to annual temperature averaged over China.[2] Given the stratification, $\sigma_h^2$ in (1) represents the within stratum (or within group) variation in stratum $h$ ($h = 1, .., L$), and $\sigma_B^2$ in (2) represents the between strata (or between groups) variation.

We don't assume that stratification is known. Therefore, we may also need to estimate stratification in our method. For a given stratification. Equations (1) and (2) can be equivalently expressed as:

$$\mathbf{y} = \mathbf{X}\boldsymbol{\beta} + \mathbf{e}, \mathbf{e} \sim N(\mathbf{0}, \sigma^2 \mathbf{I}) \quad (3)$$

where $\mathbf{y} = (\mathbf{y}_1, \cdots, \mathbf{y}_L)^T$ and $\mathbf{y}_h = (y_{h1}, \cdots, y_{hN_h})^T$. $\mathbf{X} = \text{diag}(\mathbf{1}_{N1}, \cdots, \mathbf{1}_{NL})$, $\mathbf{1}_{Nh}$ is the $N_h$-dimensional

---

[2] Note that Goldstein (2011, Eq(2.2) in p17, p19) assumed $\varepsilon_{hi} \sim N(0, \sigma_e^2)$, which is independent of the stratum $h$.

column vector with all of its components equal to 1. $\boldsymbol{\beta} = (M_1, \ldots, M_L)^T$, and $\boldsymbol{e} = (\boldsymbol{e}_1, \cdots, \boldsymbol{e}_L)^T$ with $\boldsymbol{e}_h = (e_{h1}, \cdots, e_{hN_h})^T$. **X** partitions **y** into $L$ strata in which stratum $h$ is of size $N_h$ and an element vector $\mathbf{y}_h$ and a mean $\mu_h$. We have (Gujiarati and Porter 2009, p74; p857):

$$SST = SSB + SSW \tag{4a}$$

$$SST \triangleq \Sigma^L_{h=1}\Sigma^{Nh}_{i=1}(y_{hi} - \bar{y})^2 = \mathbf{y}^T(\mathbf{I} - \mathbf{B})\mathbf{y} \tag{4b}$$

$$SSB \triangleq \Sigma^L_{h=1}\Sigma^{Nh}_{i=1}(\bar{y}_h - \bar{y})^2 = \mathbf{y}^T(\mathbf{A} - \mathbf{B})\mathbf{y} \tag{4c}$$

$$SSW \triangleq \Sigma^L_{h=1}\Sigma^{Nh}_{i=1}(y_{hi} - \bar{y}_h)^2 = \mathbf{y}^T(\mathbf{I} - \mathbf{A})\mathbf{y} \tag{4d}$$

where $\bar{y} = \Sigma^L_{h=1}\Sigma^{Nh}_{i=1}y_{hi}/N$, $\bar{y}_h = \Sigma^{Nh}_{i=1}y_{hi}/N_h$, $\mathbf{A} = \mathbf{X}(\mathbf{X}^T\mathbf{X})^{-1}\mathbf{X}^T$, $\mathbf{B} = \mathbf{1}(\mathbf{1}^T\mathbf{1})^{-1}\mathbf{1}^T$, and **1** is the $N$-dimensional vector with all components equal to 1. $SST$ denotes total variance; $SSW$ denotes within strata variance and $SSB$ denotes between strata variance. Note that (4b), (4c), and (4d) can only be derived based on a given stratification. To identify the true stratification, we need to combine them together. This motivates us to propose the statistic.

*3.2 The q-statistic: a measure of SSHy*

A measure of SSHy can be based on the ratio of SSW to SST. Wang et al (2016) define such a measure which they refer to as the *q*-statistic where

$$q(L, P) = 1 - \frac{\mathbf{y}^T(\mathbf{I}-\mathbf{A})\mathbf{y}}{\mathbf{y}^T(\mathbf{I}-\mathbf{B})\mathbf{y}} \tag{4}$$

We here make explicit that the *q*-statistic is a function of the number of strata ($L$) and the form of the particular partition ($P$) since there are many partitions that can produce $L$ strata. An optimized stratification can be founded by

$$O(L, P) = \text{argmax}_{(L, P)}\{q(L, P)\} \tag{4e}$$

The essential difference between the R-square or the interclass correlation coefficient (ICC) (Donner and Koval 1980) and our *q*-statistic given by (4) is that R-square or ICC assumes that the partition is given but we don't make such assumption. That leads to *q*-statistic following non-central $F$ PDF (see the next paragraph) while ICC following $F$ PDF (Snijders and Bosker RJ 2011, p46). *q* statistic is used to detect SSH and to attribute SSH without linearity assumption, that are another different from R2 and ICC and multilevel modelling (MLM). In Figure 4 we illustrate the *q*-statistic. Note that for emphasis we have shown stratum $h = 1$ appearing in two distinct geographical areas. The stratum label refers to its classification (e.g. land use type) not its spatial location. The stratification of a variable $Y$ can be partitioned either by $Y$ itself or by a suspected explanatory variable $X$ (section 4.3), depending on the purpose of the study.

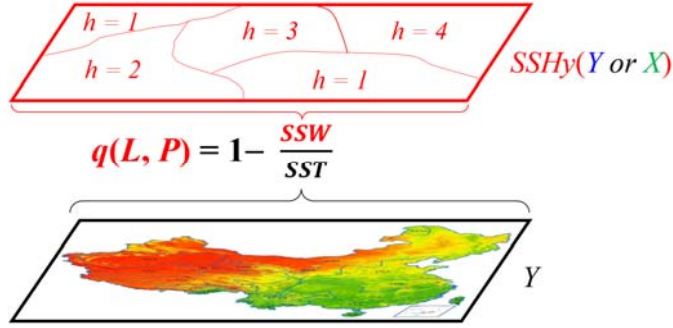

**Figure 4**. Illustrating the $q$-statistic

The $q$-statistic, for any given $L$ and $P$, takes a value in the interval [0, 1], where 0 reflects stratified heterogeneity, which implies that each of the $L$ strata have the same degree of (internal) heterogeneity as found across the whole map (Fig 3a). A value of the $q$-statistic equal to 1 indicates complete within-stratum homogeneity (all data values within a stratum are the same), which implies that the heterogeneity observed across the whole map is due to the differences between the $L$ strata for the given partition (Fig 3d). Given $u_1, \ldots, u_L$, the exact probability density function of the $q$-statistic is a non-central $F$ function:

$$F = \frac{N-L}{L-1} \frac{q}{1-q} \sim F(L-1, N-L; \lambda) \tag{5}$$

$$\lambda = \frac{1}{\sigma^2}[\sum_{h=1}^{L} N_h u_h^2 - \frac{1}{N}(\sum_{h=1}^{L} N_h)^2] \tag{6}$$

where $N$, $L$, $q$, $\sigma^2$ and $N_h$ all have the same meanings as in the preceding equations.[3] Mathematical result (5) and (6) can be obtained as follows. From (1) and (2), we have

$$y_{hi} - \bar{y}_h = e_{hi} - \frac{1}{N_h}\sum^{N_h}_{i=1} e_{hi} = e_{hi} - \bar{e}_h$$

thus

$$SSW = \Sigma^L_{h=1}\Sigma^{N_h}_{i=1}(e_{hi} - \bar{e}_h)^2 = \mathbf{e}^T(\mathbf{I-A})\mathbf{e}.$$

Note that A - B is an orthogonal projection matrix and tr(I – A) = $N - L$. We have

$$SSW \sim \sigma^2 \chi^2_{N-L}$$

Given $\mathbf{u} = (u_1, \ldots, u_L)^T$, we can treat them as constants. Then,

---

[3] This result holds for other forms of stratification such as in time if data values are temporal.

$$\sqrt{N_h}(\bar{y}_h - \bar{y}) = \sqrt{N_h}(u_h - \frac{\sum_{i=1}^{N} N_i u_i}{N}) + \sqrt{N_h}(\frac{1}{N_L}\sum_{i=1}^{N_h} e_{hi} - \frac{1}{N}\sum_{h=1}^{L}\sum_{i=1}^{Nh} e_{hi})$$

Thus, we have

$$\lambda = \frac{1}{\sigma^2}\sum_{h=1}^{L}[\sqrt{N_h}(u_h - \frac{\sum_{i=1}^{N} N_i u_i}{N})]^2 = \frac{1}{\sigma^2}[\sum_{h=1}^{L} N_h(u_h - \frac{\sum_{i=1}^{N} N_i u_i}{N})^2]$$

$$= \frac{1}{\sigma^2}[\sum_{h=1}^{L} N_h u_h^2 - \sum_{h=1}^{L} N_h(\frac{\sum_{i=1}^{N} N_i u_i}{N})^2]$$

$$= \frac{1}{\sigma^2}[\sum_{h=1}^{L} N_h u_h^2 - \frac{1}{N}(\sum_{h=1}^{L} N_h u_h)^2]$$

*3.3 Evaluation of different stratifications*

In section 1 we have noted the importance of being able to identify homogeneous zones. This is not a new challenge. The identification of such zones lies at the heart of traditional regional geography as well as contemporary data analysis practice concerned with boundary delineation (e.g. wombling – see Womble 1951) and various forms of regional classification or "region-building" (Longley et al. 2005, p135; Dutilleul 2011, p.20-21). As the stratification is unknown, we want to find the stratification with minimum within group variation and maximum between group variation. For a review of spatial and non-spatial regional classification methods, such as the K-means (MacQueen 1967), see for example (Haining, 2003, p.199-201). In the case of spatial classifiers (Haining, 2003, p.201-206), the objective function in these methods is usually constructed by combining a homogeneity function for within-group variances and a spatial compactness function for their coordinate locations. The *q*-statistic proposed in this article can be treated as the first of these functions. The advantage of the *q*-statistic as we shall see, is that it is possible to connect its null distribution with a standard, well-known, distribution such that we can easily derive its *p*-value. The *q*-statistic can be used to make an empirical evaluation of different stratifications (varying *L* and/or varying the partition for the same value of *L*) to see which yields the largest value of the statistic. Take the case where *L* is fixed (we are confident about the number of strata) and we want to compare two partitions *P1* and *P2* that yield *L* strata. Compute:

$$Q(P1, P2) = q(L; P1) - q(L; P2) \qquad (7)$$

If $Q(P1, P2) > 0$ ($< 0$) then *P1* (*P2*) yields a more homogeneous intra-stratum partition than *P2* (*P1*). Because $Q(P1, P2) = \frac{SSB1 - SSB2}{SST} = \frac{D(P1,P2)}{SST}$, the statistical significance of the difference between the two partitions can be tested by *D(P1, P2)*. If *P1* is the true stratification, then it can be

shown that:

$$D(P1, P2) \sim^{\text{approximate}} N(E(D), V(D)) \quad (8)$$

where

$$E(D) = \sigma^2 \text{tr}(\mathbf{A}1 - \mathbf{A}2) + \boldsymbol{\beta}^T \mathbf{X}_1^T (\mathbf{A}1 - \mathbf{A}2) \mathbf{X}_1 \boldsymbol{\beta} \quad (9)$$

$$V(D) = 2\text{tr}(\mathbf{A}1 - \mathbf{A}2)^2 + 4\boldsymbol{\beta}^T \mathbf{X}_1^T (\mathbf{A}1 - \mathbf{A}2)^2 \mathbf{X}_1 \boldsymbol{\beta} \quad (10)$$

where $E$ and $V$ stand for expectation and variance, respectively; $\mathbf{A}1$ and $\mathbf{A}2$ are $\mathbf{A}$ for $P1$ and $P2$, respectively; $\mathbf{X}_1$ is $\mathbf{X}$ for one of the partitions, $P1$, say. Whilst there may be many partitions generating $L$ strata, in practice the number of partitions that will justify comparison should be much fewer in number.

The individual elements of $q(L; P1)$, $SSW_h$, that is the $L$ terms in the numerator of the second term in the definition of $q(L; P1)$ in (4), can be compared to see which of the various strata display most (intra-stratum) heterogeneity. The strata making the largest contribution to the numerator might be candidates for further partitioning – that is increasing $L$. However, comparing partitions involving different numbers of strata raises another problem, namely the need to include a penalty to prevent over-stratification. The argument is similar to that encountered when selecting independent variables in a regression model where Akaike's Information Criterion (AIC) is used to compare different models allowing for differences in model complexity. A model, A, with more independent variables than another, B, where the independent variables in B form a subset of those in A, will fit the data better but is more complex and this needs to be allowed for when comparing model fits – in this case two stratifications with a different number of strata. AIC is penalized estimation based on minimized Kullback-Leibler information between two probability density functions and might be used to compare two stratifications that differ in terms of $L$ (Akaike 1974).

**4 Inference under Spatially Stratified Heterogeneity**

In this section, we consider examples of statistical analysis when spatially stratified heterogeneity is present in a dataset.

*4.1 Statistics within homogeneous strata*

Once a partition into homogeneous strata has been constructed, if observations are independent[4],

---

[4] For example, sample points are sufficiently far apart that observed values are independent. In a

conventional statistics can then be applied to draw inferences about intra-strata properties as the example in section 2 illustrates (Xu et al. 2011, see Figures 2b and 2c). If, on the other hand, intra-stratum observations are spatially autocorrelated, then spatial statistical techniques are needed to draw inferences about intra-strata properties. Since many of these statistics depend on the specification of a weights (or connectivity) matrix $W$ (Haining and Li 2020, chapter 4) for spatial relationships between the observations, a question arises as to how to treat those observations close to the boundary of any stratum. In the interest of improving statistical precision (particularly if a stratum has relatively few observations), might it be appropriate to include some observations, suitably weighted, from the "other side" of the stratum boundary? The justification for such an approach is because some segments of a stratum boundary may be more akin to a "zone of transition" even if other boundary segments display abrupt change. One way to try to resolve such a question is to *estimate* the non-zero entries in the $W$ matrix (Haining and Li 2020, 4.10 and 8.4). Lee and Mitchell (2013) describe a method using locally adaptive spatial smoothing. The basis of their approach is to treat the non-zero elements in the $W$ matrix as random variables (that may be reset to zero under a decision rule) whilst the zero elements of the weights matrix remain as zeroes.

A number of techniques are proposed to construct maps of the spatial distribution of some SSH *attribute* (e.g. annual air temperature by climatic zones) on the basis of a sample of spatially autocorrelated observations. MSN (means of surface with non-homogeneity), B-SHADE (biased sentinel hospital area disease estimator) and SPA (single point area) estimators combine Kriging and stratified sampling to make inferences that are best linear unbiased (BLUE). MSN is applicable when all strata have samples (Wang et al. 2009; Hu et al. 2011; Gao et al. 2019; Wang et al. 2019), the estimator reduces to Kriging if SSHy is absent and to the Sandwich estimator (Wang et al. 2013b) if SAC is absent. When some strata have no observations, B-SHADE uses the ratio between a sample and the population. This ratio may be estimated using a covariate. For example, the ratio in the early years when stations are few in number can be estimated by the observations from present-day meteorological stations. They are greater in number, to adjust for sample bias (Wang et al. 2011; Hu et al. 2013; Xu et al. 2013, 2018). B-SHADE reduces to MSN if all strata have samples. When only a single sample unit is available, SPA estimates an area mean using a prior relationship

---

geostatistical analysis this occurs when the inter-sample distance exceeds the range.

that has been identified between the target variable (PM2.5, for example) and a covariate (PM10, for example) which has been observed in all strata (Wang et al. 2013a). MSN, B-SHADE and SPA estimates are obtained by a simple matrix transform.

If interest centres on the spatial distribution of some *parameter*, Haining and Li (2020, chapters 7 and 8) describe a number of Bayesian Hierarchical Models (BHMs) with spatial dependence that yield estimates of a heterogeneous parameter. They illustrate the application of these methods to samples of household income data for Newcastle-upon-Tyne, England at the middle super-output area (MSOA) level. An MSOA is a small area that is used for reporting UK Census data and which here we treat as a single stratum. The sample is collected from a national survey so that not all MSOAs (strata) have data and many have only a small number of observations. The samples within each stratum are assumed to be independent and identically distributed. It is the set of MSOA-level parameter values that are spatially autocorrelated. The application involves different autoregressive models for capturing spatial autocorrelation in the spatial distribution of the parameter of interest (average household income at the MSOA level). The chosen model, specified in the BHM's prior model for the parameter of interest, leads to information sharing across the MSOAs. BHMs are fitted using Markov chain Monte Carlo (MCMC) simulation. Here as in the cases of the other methods described above, additional covariates can be included in the model to improve estimates.

*4.2 Estimation from one heterogeneous spatial framework to another*

The term spatial framework refers to a set of spatial units or supports. China composed of 34 provinces and China composed of a set of climate zones are examples of two spatial frameworks. Areal interpolation is the term used to describe the process of transferring data from one spatial framework ("the source") to a new spatial framework ("the target" or "reporting" framework) – see for example Goodchild et al. 1993 and Haining 2003 p.131-8 for an overview. Wang et al. 2013b describe what they term "the sandwich method" to transfer data values from a source to a target framework which is appropriate for an SSH population (Fig 5). The methodology provides estimates for each target zone together with an estimate of error variance. First the SSH population is stratified into homogeneous strata (Wang et al. 2010a) and mean and variance estimates are obtained for each of the source strata ({*h*} in Fig 5). Next, the SSH population of source zones is overlaid on the target zone framework ({*r*} in Fig 5). The estimates for the target zones are obtained from the source zones in proportion to the degree of overlap that occurs. Any individual source zone may contribute to the estimate of more than one target zone. In that sense the sandwich approach "borrows strength" from all the source strata that overlap any particular target stratum and depending on the geographical extent and configuration of the source zones relative to the target zones, this process of borrowing strength may not be limited to just nearby areas. Providing there are sample values in each source zone the methodology can be applied and does not rely on their being sample data in a target unit. Whilst kriging outperforms sandwich sampling if neighbouring values are more similar than those within an SSH stratum, sandwich sampling works best when spatial autocorrelation is either absent or weak (Liu et al. 2018) (A population $\Re$ is composed of strata $\{h = 1, ., L\}$ and reporting units $\{r\}$. The green shaded area in $\{h\}$ takes the example of stratum $h = 4$. The red transparent prism between $\{h\}$ and $\{r\}$ illustrates information flowing from $\{h\}$ to $\{r\}$; $\bar{y}$ and $V\bar{y}$ stand for mean and variance of the attribute $y$, respectively; $n$ and $N$ stand for the number of sample units and all units in a stratum, respectively. Subscript $h$ stands for stratum $h$, $hi$ stands for the $i$-th sample unit in stratum $h$; subscript $rh$ stands for a unit formed by the intersection between two units $r$ and $h$.)

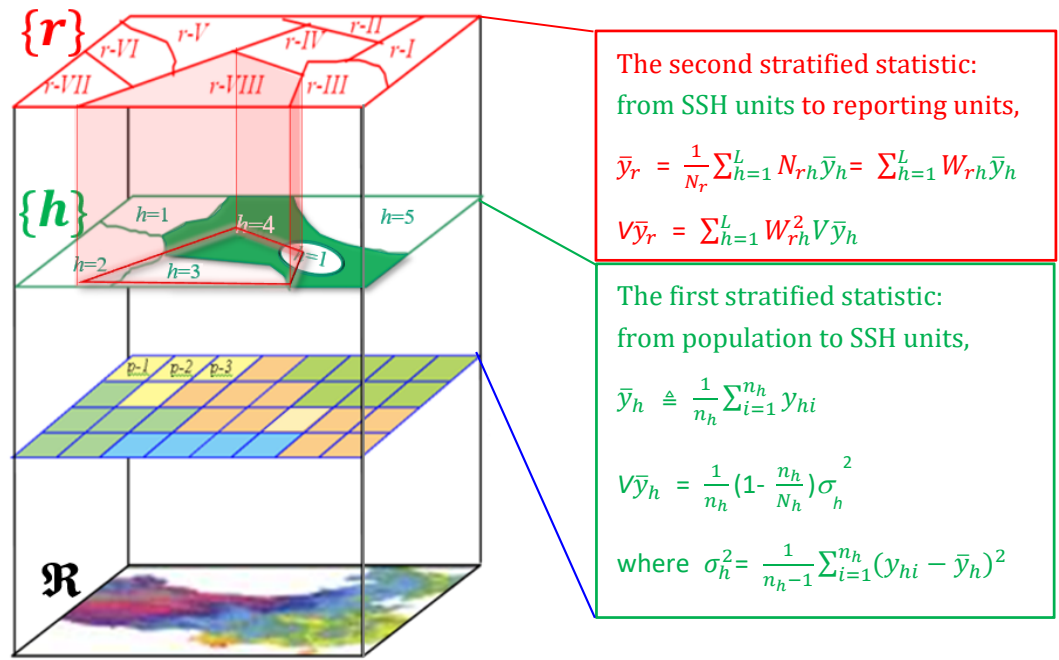

**Figure 5**. Illustrating the sandwich estimator when the population is SSH

*4.3 Using the q statistic to study associations between variables.*

If *X* is linearly related to *Y*, their spatial distributions will correspond. Pearson's correlation coefficient (*r*) can be used to measure and to test for the significance of the relationship. However, if the relationship is not linear, Pearson's statistic does not provide a satisfactory measure of association. When examining associations over large geographical areas which can be partitioned into several strata, it may be of interest to employ a more flexible indicator of association as suggested by the example shown in Figure 2 and Yin et al. (2019).

Our axiom is that if *X* are related *Y* (perhaps causally), their spatial patterns would appear consistence. A version of the statistic $q(L; P)$ also can be used to examine the consistence, as hinted at in Figure 4. In Figure 4 we note that the partition, *P*, can be specified in terms of the observed values of *Y* which is the variable to be tested for stratified heterogeneity, or it could be specified in terms of another variable, say *X*. To make it clear which variable is being used to construct the partition we could write for example $q(L; P_z)$ when the partition is on some variable *Z*. This variable, *Z* might be either a categorical variable such as landuse type or a quantitative variable having the property to it that the values of *Z* within any of the *Z*-defined strata, are similar, but differ between strata. Suppose that we are able to specify a partition, $P_x$, of a quantitative variable *X*. The mean value of *X* within a single stratum defines the level of *X* in that stratum and all the observed values of *X* in that same stratum are similar thereby giving rise to a large value of the *q*-statistic calculated on the *X* values. The mean value will be different in other strata but there will be little variation in the *X* values within those strata. If there is an association between *Y* and *X* across the strata then the value of the statistic, call it $q(L; P_x)$, calculated on *Y* will tend to be large (closer to 1). On the other hand, if there is no association between *Y* and *X*, the value of $q(L; P_x)$, calculated on *Y* will tend to be small (closer to 0). This is because if there is an association between *X* and *Y* then, within the strata, we should expect homogeneity in *X* to be associated with homogeneity in *Y* (Li et al. 2020). In other words, if there is an association between *X* and *Y* in a study area, then their spatial patterns, depicted by stratifications $(L; P)$, tend to be coupled. The degree of the consistence can be measured by $q(L; P_x)$ of *y*.

The consistence between two long non-monotonous time series is more difficult to appear than a simple linear correlation between the two variables at a pair of points (Figure 6ab), in addition of the irreversibility of time arrow, implies the causation between agriculture production and the north hemisphere temperature (Zhang et al. 2007). Experiments illustrate that a spatial population is more complicated than an aspatial population (MacEachren et al. 1982), and a choroplethic map is more complicated that an isopleth map (Monmonier 1974), ceteris paribus. By the same philosophy, the consistence between two complicated spatial patterns needs not only the coincidence in places but also identical shapes of the two variables, so is more difficult to appear than the correlation between two pairs of points (Figure 6ac), implying direct or indirect causation between the two spatial variables (Figure 6c, Figure 4), and called as SSH causality.

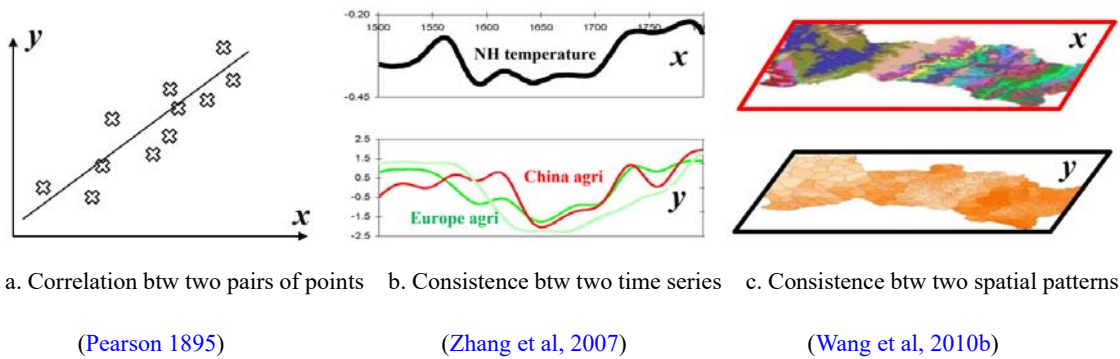

a. Correlation btw two pairs of points    b. Consistence btw two time series    c. Consistence btw two spatial patterns

(Pearson 1895)            (Zhang et al, 2007)            (Wang et al, 2010b)

**Figure 6**. Association between two points' pairs, two time series, and two spatial patterns: from correlation to causation

If $X$ is a quantitative variable, then a bivariate plot of the means of $X$ against the corresponding means of $Y$ by strata will indicate the form of any association – which need not be linear. Note that the two values of $q(L; Px)$ one calculated on $Y$ the other on $X$, but using $Px$ in both cases, is required to determine the extent to which the two variables are homogeneous within the strata. Figure 7 decomposes the components of the $q$ statistic when the aim is to compare two variables ($X$ and $Y$). We assume that the stratification of the map has been based on the spatial variation in the variable $X$ (not $Y$). Each circle on the scatterplot refers to a single $X$-defined stratum. The *centre* of any circle on the scatterplot is defined by the mean values of $Y$ and $X$ within that stratum, respectively. The *size* of any circle is proportional to the size of the stratum (in terms of population size or areal extent) in order to give visual weight to larger strata. However the *shading* of any circle (for stratum $h$) is based on $q_h = 1 - [\sigma_h^2/\sigma^2]$ *calculated on the variable Y*. The darker the shading the smaller the within stratum variance of $Y$ (the within stratum variance on $X$ is small by construction). We might be able to refine this plot further. If any stratum $h$ comprises several discrete geographical areas (see

Figure 4), then rather than one circle for stratum *h* the circle could be disaggregated with one for each discrete area. This might be desirable to do where, for example, the areas are large and widely scattered geographically. It might be undesirable to do this if each discrete geographical area is small giving rise to a "small number" problem (see for example Haining and Li 2020, p.81).

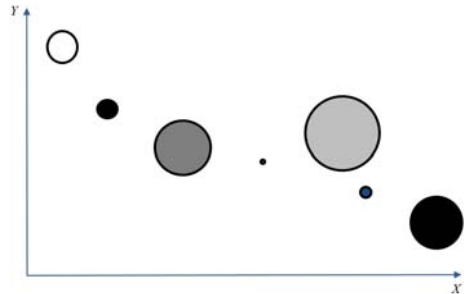

**Figure 7**. *q*-statistic scatterplot suggesting, descriptively, a weak, inverse relationship between *X* and *Y* at the scale of 7 aggregate strata of varying size. There may be a case for disaggregating the large circle, third from the right, if it comprises several discrete geographical areas. The same comment might apply to the two other larger circles – first from right and third from left.

5. **Empirical Studies**

In this section we review two empirical studies in geographical (or spatial) epidemiology where the presence of spatially stratified heterogeneity raises methodological issues.

*5.1 Obtaining a map of breast cancer incidence using sample data: comparing sandwich sampling and kriging. (China National Cancer Centre 2019)*

Breast cancer mortality rates in the over 2700 counties of China in 1992, were collected from China's Centre for Disease Control (CDC) (Figure 8a). Moran's I test for spatial autocorrelation (using the four nearest counties as neighbours when specifying the *W* matrix) gave a value of 0.195 ($p = 0.001$). We calculated *q* for 8 different partitions ($L = 3, 4, …, 10$ strata) of the breast cancer mortality data and the partition is based on selecting a large value of *q* whilst restricting the number of partitions (*L*). We chose the stratification ($L = 5$) which gave rise to the first significant value of *q* at the 1% level ($q = 0.6381$ with $p < 0.01$) whilst "penalizing" for increasing values of *L* (see Table 1). As we move from *L*=4 to *L*=5 *q* increases by 0.74 (from 0.564 to 0.638) and from a *p*-value of 0.029 (not significant at the 1% level) to a *p*-value of 0.007 (significant at the 1% level). At the next level of stratification from $L=5$ to $L=6$ the improvement in $L$ (which by definition must occur) is less than occurred when moving from 4 to 5 strata (0.061 compared to 0.074) indicating a decrease in the size of the improvement relative to the improvement when moving from $L=4$ to $L=5$. We are implicitly invoking a criterion of simplicity when trying to fix $L$ – a problem we will discuss again in section 6. These two statistics (Moran's $I$ and the $q$-statistic) indicate that the breast cancer mortality data display both spatial autocorrelation and SSHy at the county support.

**Table 1. $q$-statistic calculated for breast cancer mortality rates in China in 1992**

|  | \multicolumn{8}{c}{$L$} | | | | | | | |
|---|---|---|---|---|---|---|---|---|
|  | 3 | 4 | 5 | 6 | 7 | 8 | 9 | 10 |
| $q$ statistic | 0.400 | 0.564 | 0.638 | 0.699 | 0.733 | 0.774 | 0.818 | 0.823 |
| $p$ value | 0.024 | 0.029 | 0.007 | 0.000 | 0.000 | 0.000 | 0.000 | 0.000 |

The SSH based sandwich method used the mortality data from 64 sample counties (see Figure 8a) to construct a county-level map of breast cancer mortality for all the counties of China in 1992 (Figure 8c). The results obtained by this method and by Kriging (Figure 8b) were compared with the known mortality rates from the CDC database. The leave-one-out validation $R^2$ derived from sandwich mapping is 0.705, whereas for Kriging the value is 0.532. In this instance therefore and by the $R^2$ criterion, sandwich mapping outperforms Kriging. Aside from the many different reporting units that could have been presented derived from the original sample, the sandwich method has the particular merit of invoking simple modelling assumptions coupled with full utilization of the sample data through stratification rather than using only nearest neighbour data values as in the case of Kriging. Furthermore, whilst kriging, because it is a spatial autocorrelation based interpolation method has to address boundary effects, which is not the case with the sandwich method. The evidence from this example suggests that the sandwich method performs well, and seems to perform better than kriging, when a variable exhibits SSHy. This clearly needs further investigation to determine whether this holds up more generally and particularly when methods of kriging, developed for heterogeneous surfaces, are implemented.

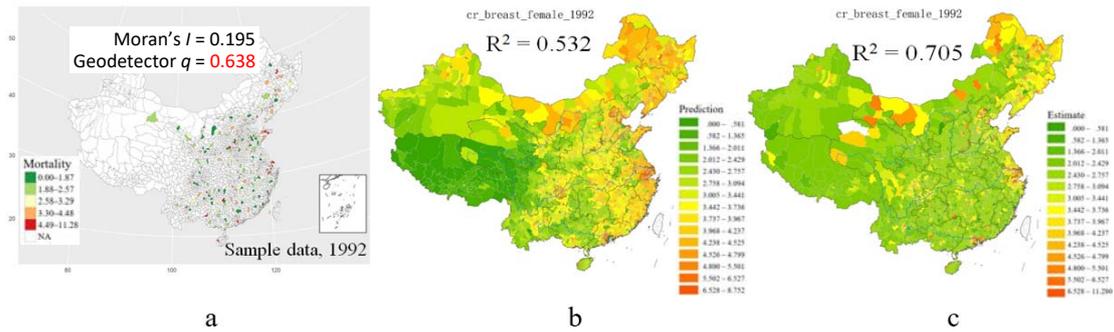

**Figure 8**. Interpolation of breast cancer mortality sample data (a) using SAC based Kriging (b) and SSH based sandwich (c) methods

*5.2 Geodetector: analysing the distribution of cases of neural tube defects in Heshun County* (Wang et al. 2010b)

We summarize findings from a pilot project investigating the factors associated with the geographical distribution of cases of Neural Tube Defects (NTD) in Heshun County, Shanxi Province (Wang et al. 2010b), which is an area with one of the highest NTD incidence rates in China. The factors ($X$) believed to be responsible for variation in NTD incidence ($Y$) include the physical environment, man-made pollution and nutrition. The factors ($X$) are measured at both a nominal level and a quantitative level and exhibit SSHy (see Figure 9). The *q*-statistic (Figure 4) was used to examine the spatial association between NTD incidence and the suspected factors ($X$). The procedure for examining the association between a $Y$ and any particular $X$ is as follows. First, the map for the selected $X$ is stratified. Second, the map for $Y$ is overlaid on the stratified map of the chosen $X$. Finally using the data on $Y$ calculate $q(L, Px)$ (see section 4.3). Software to calculate $q$ is available at www.geodetector.cn.

Figure 10 presents the *q*-statistic scatterplot between NTD incidence ($Y$) and elevation ($X$), a proxy for one of the suspected determinants. The scatterplot visualizes the SSH information of the data. For example, NTD incidences at elevation 1300m is high (6.9%) but very variable as indicated by the white colour of the circle moreover the stratum area to which these data refer is small (as indicated by the size of the circle). The four dark circles on the plot indicate elevations where NTD incidences are not highly variable but there is no evidence of a simple linear relationship between NTD incidence and elevation.

Table 2 presents the *q*-statistic between NTD incidence (*Y*) and three of the suspected determinants (*Xs*). The study found that some environmental factors (watershed type and elevation but not soil type) were found to be significantly associated with variation in NTD occurrences in the region. Usually water quality and the geological chemical environment are more similar within watersheds than between watersheds. Non-environmental factors (not reported here) were of secondary importance. These findings were helpful for identifying what courses of action would be most appropriate for disease intervention in the region (see Wang et al. 2010b).

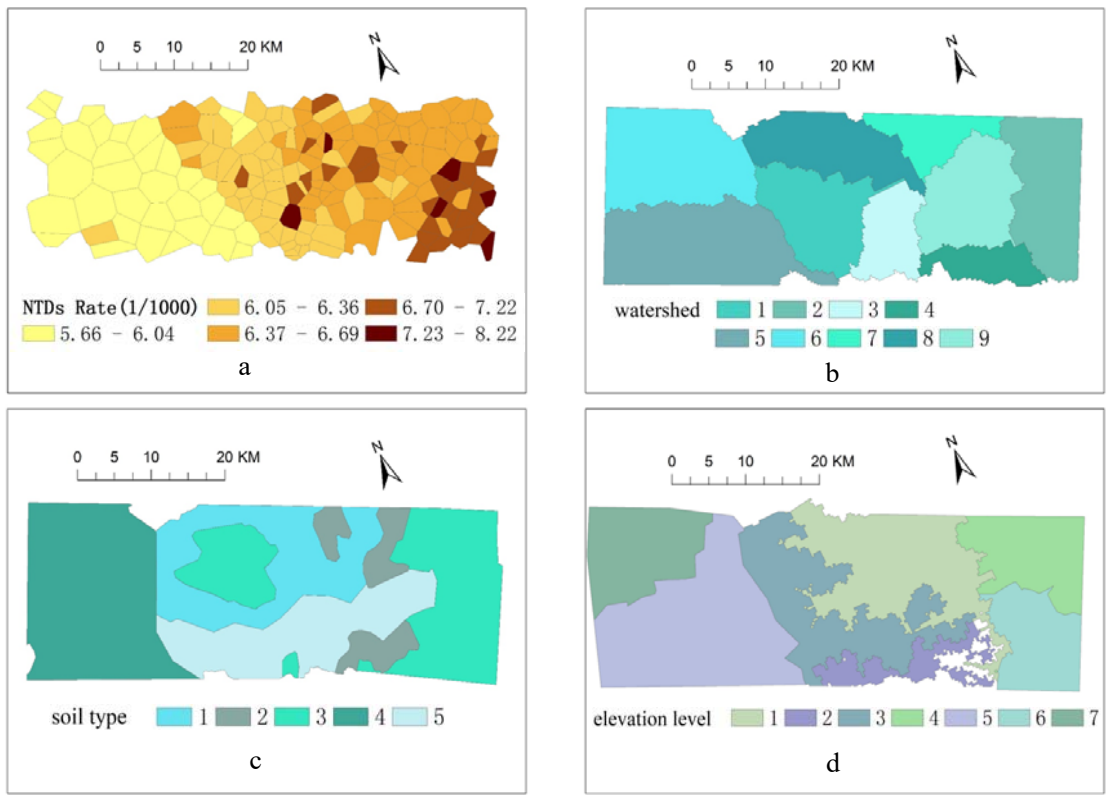

**Figure 9**. Example data of NTD incidence (*Y*) (top left figure) and suspected factors (*X*) (b. watershed; c. soil type; and d. elevation level) (adapted from Wang et al. 2010b)

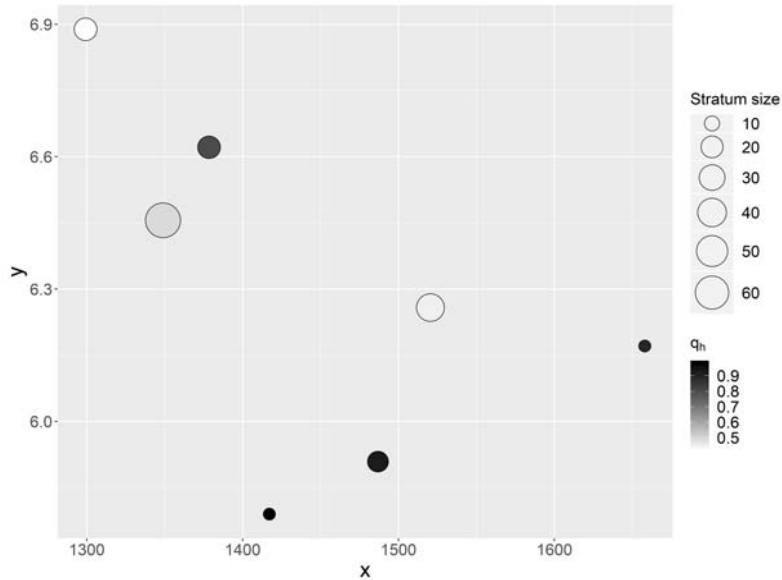

**Figure 10**. *q*-statistic scatterplot between NTD incidence (*Y*) and elevation (*X*)

**Table 2. *q*-statistic calculated for NTD incidence using strata constructed for different suspected determinants**

|  | watershed | elevation | soil type |
|---|---|---|---|
| *q* statistic | 0.64 | 0.61 | 0.39 |
| *p* value | 0.00 | 0.04 | 0.36 |

In both of the examples and in several of the earlier sections (e.g. 2.2) statistical results depend on the scale (size) of, and partitioning associated with, the reporting units (referred to as the modifiable areal units problem, or MAUP, Openshaw (1984); Ge et al. 2019).  This is an issue endemic within spatial analysis when working with areal units.  Typically, the methods described here do not appear to introduce any additional, significant scale effects because the reporting units themselves are unaffected (with the possible exception of some uses of sandwich sampling where some source zones might be divided up and distributed over two or more target zones). However, the creation of homogenous subsets does add another layer of the "partition effect". This is why it is important to be able to detect and justify the presence of spatial heterogeneity and to implement the methodology to identify homogeneous subsets as described in section 3.3 and discussed further below.

## 6. Discussion and Conclusions

SSHy is often an important characteristic of spatial data especially when data cover a large geographical area or where the data are in high resolution. The problems caused by SSHy to statistics that assume homogeneity can be resolved or at least reduced if the structure of the heterogeneity can be identified such that the map can be partitioned into homogeneous subpopulations. Conventional (iid) statistical methods can be used in each stratum if observations are independent, spatial statistical methods should be used if the observations are spatially autocorrelated. If the structure of the SSHy can be identified, then it can also be used to design sampling schemes that will help in drawing a representative sample that provides coverage of the different strata and reduce estimator bias.

Identifying SSHy offers an opportunity for spatial interpolation when spatial autocorrelation is absent or weak. Besides providing a measure of SSHy, the $q$ statistic can be used to explore the association between the spatial patterns of two variables, as was discussed in section 4.3. Table 3 summarizes tools that are often used for data showing different states (presence/absence) of SSHy and SAC.

Table 3. Examples of techniques when working with spatially stratified heterogeneous data

| The statistical characteristics of geographical variables | | Spatially stratified heterogeneity (SSHy) (measured by $q$ statistic) | |
|---|---|---|---|
| | | Absent | Present |
| Spatial autocorrelation (SAC) | Absent | Classical (iid) statistics | Spatial classification methods; wombling; Areal interpolation; Sandwich estimation |
| | Present | Spatial statistics applied to identically distributed (id) data | MSN (stratified sample); B-SHADE (biased sample); SPA (single point sample); Hierarchical modelling |

Although in certain research contexts SSHy is endemic and numerous algorithms have been developed for spatial classification, SSHy has not received the attention that other special characteristics of spatial data have received. The development of statistics for SSHy, will add to the toolbox of exploratory spatial analysis techniques, alongside techniques for exploring data that are spatially autocorrelated (Anselin 1995).

Cressie (1993, p.25) remarked: "Whether one chooses to model the spatial variation through the (non-stochastic) mean structure (called large scale variation….) or the stochastic dependence structure (called small scale variation) depends on the underlying scientific problem, and can sometimes be simply a trade-off between model fit and parsimony in model description. *What is one person's (spatial) covariance structure may be another person's mean structure.*" This remark is relevant to the relationship between the characteristics of heterogeneity and spatial autocorrelation and how each might be handled in the course of data analysis (Atkinson 2000). For example, the behavior of LISA's, Getis-Ord statistics and Geographically Weighted Regression, all exploratory statistics, may be reflecting either local spatial autocorrelation (Anselin 1995; Sui 2006, p494) or spatial heterogeneity (Goodchild and Haining 2004; Anselin 2006). The semi-variogram, used in geostatistics, underpins kriging. It is an interpolation method based on the spatial autocorrelation properties of a set of data. The method can also be employed to identify spatial heterogeneity (see, Issaks and Srivastava 1989, p223-224 and Fig.9.5; p292 and Fig.12.2 also discuss the use of the correlogram in kriging). The two concepts, spatial heterogeneity and spatial autocorrelation, do not reflect distinct and separate features of spatial data. To paraphrase Cressie, "one person's heterogeneity in the mean may be another person's local or global spatial autocorrelation". That said, one of the central points of this paper has been to argue that spatially stratified heterogeneity may, as in the examples presented here, fit better with the scientific problem and provide a satisfactory trade-off between model fit and model complexity.

As a new tool to analyse SSHy, we believe two issues related to the $q$ statistic need further work. First, the value of $q(L, P)$ depends upon both the number of strata and the spatial structure of the stratification. In some circumstances there may be a large number of plausible stratifications (in terms of both $L$ and $P$) and efficient methods are needed to compare them. Closely related to this point, the decision on the number of strata to employ involves a trade-off between complexity (the number of strata) and the level of intra-stratum homogeneity as discussed in section 5. (How can we be sure that we have chosen the best $L$ and $P$? Should it just be a statistical decision? how much substantive knowledge should be drawn upon to make the final choice and how sensitive are our results to the chosen $L$ and $P$?) In keeping with other forms of statistical decision making the AIC statistic may provide a way to formalize this trade off as suggested in section 3.3 but more work is needed to assess this because AIC is not applicable directly. The reason is that the pdf which AIC is based on is not on a one to one mapping with the spatial distribution which the $q$ statistic is based on. For example, different spatial distributions (stratifications) may share the same pdf.

Second, whilst spatial autocorrelation and SSHy are two important characteristics of spatial data, the relationships between the two and the influence of each upon the other in the conduct of spatial data analysis, are in need of further investigation. Notwithstanding these concerns there are a few rules of thumb that should be helpful in choosing (*L, P*) as we illustrated in 5.1.

**Acknowledgement**

This study was supported by the National Natural Science Foundation of China (Nos. 42071375; 41531179); MOST (2022YFC3600800); National Social Science Foundation of China (No. 21&ZD186). We appreciate the assistance given by Ge Y, Liao YL and Li LF in this study.

**References**

Aiello, F., Ricotta, F. 2016. Firm heterogeneity in productivity across Europe: evidence from multilevel models. *Economics of Innovation and New Technology* 25(1): 57-89.

Akaike, H. 1974. A new look at the statistical model identification. *IEEE Transactions on Automatic Control* AC-19: 716-723.

Anselin, L. 1995. Local indicators of spatial association – LISA. *Geographical Analysis* 27(2): 93–115.

Anselin, L. 2006. Spatial Heterogeneity, In B. Warff (Ed.), *Encyclopedia of Human Geography*. Thousand Oaks, CA, Sage Publications, p.452-453.

Atkinson, P., Tate, N. 2000. Spatial scale problems and geostatistical solutions: a review. *Professional Geography* 52(4): 607-623.

Bradley VC, Kuriwaki S, Isakov M, Sejdinovic D, Meng XL, Flaxman S. 2021. Unrepresentative big surveys significantly overestimated US vaccine uptake. *Nature* 600 (7890): 695-700.

Buttle, J. M., D. M. Allen, D. Caissie, B. Davison, M. Hayashi, D. L. Peters, J. W. Pomeroy, S. Simonovic, A. St-Hilair, and P. H. Whitfield. 2016. Flood processes in Canada: Regional and special aspects. *Canadian Water Resources Journal*. DOI:10.1080/07011784.2015.1131629.

China National Cancer Centre. 2019. *Cancer Atlas in China. China*. Beijing: Sinomap Press.

Christakos, G. 1992. *Random Field Models in Earth Sciences*. CA, San Diego: Academic Press.

Cliff, A.D., J. K. Ord. 1981. *Spatial Processes: Models and Application*. London: Pion.

Cressie, N. 1993. *Statistics for Spatial Data*. New York: Wiley.

Dormann, C. F., J. M. McPherson, M. B. Araujo, R. Bivand, J. Bolliger, G. Carl, R. D. Davies, A. Hirzel, W. Jetz, W. D. Kissling, I. Kuhn, R. Ohlemuller, P. R. Peres-Neto, B. Reineking, B. Schroder, F. M. Schurr, and R. Wilson R. 2007. Methods to account for spatial autocorrelation in the analysis of species distributional data: a review. *Ecography* 30: 609-628.

Dunn, R., and A. R. Harrison. 1993. Two dimensional systematic sampling of land use. *Applied Statistics* 42: 585-601.

Dutilleul, P. R. L. 2011. S*patio-Temporal Heterogeneity: Concepts and Analysis*. Cambridge: Cambridge University Press.

Everitt, B. S., and A. Skrondal. 2010. *The Cambridge Dictionary of Statistics*. 4th Ed. Cambridge University Press.

Fazio, G., Piacentino, D. 2010. A spatial multilevel analysis of Italian SMEs' productivity. *Spatial Economic Analysis* 5(3): 299-316.

Fotheringham, A. S., C. Brunsdon, M. Charlton. 2000. *Quantitative Geography: Perspectives on Spatial Data Analysis*. London: Sage.

Gao, B. B., J. F. Wang, M. G. Hu, H. M. Fan, and K. Xu. 2015. A stratified optimization method for a multivariate marine environmental monitoring network in the Yangtze River estuary and its adjacent sea. *International Journal of Geographical Information Science* 29(8): 1332-1349.

Ge, Y., Y. Jin, A. Stein, Y. H. Chen, J. H. Wang, J. F. Wang, Q. M. Cheng, H. X. Bai, M. X. Liu, P. Atkinson. 2019. Principles and methods of scaling geospatial earth science data. *Earth-Science Review* 197: 102897.

Getis, A., and J. K. Ord. 1992. The analysis of spatial association by use of distance statistics. *Geographical Analysis* 24: 189-206 (with correction, 1993, 25, p. 276).

Goldstein, H. 2011. *Multilevel Statistical Models, 4th Edition*. Wiley.

Goodchild, M. F., L. Anselin, and U. Deichmann. 1993. A framework for the areal interpolation of socioeconomic data. *Environment and Planning A* 25: 383-297.

Goodchild, M., and R. Haining. 2004. GIS and spatial data analysis: converging perspectives. *Papers in Regional Science* 83: 363-385.

Griffith, D. A. 2003. *Spatial Autocorrelation and Spatial Filtering, Gaining Understanding Through Theory and Visualization*. Springer-Verlag, Berlin.

Gujarati, D. N., D. C. Porter. 2009. *Basic Econometrics, 5th Edition*. McGraw-Hill.

Haining, R. 2003. *Spatial Data Analysis: Theory and Practice*. Cambridge: Cambridge University Press, Cambridge.

Haining, R., and G. Q. Li. 2020. *Modelling Spatial and Spatial-Temporal Data: A Bayesian Approach*. CRC.

Heckman, J. J. 1979. Sample selection bias as a specification error. *Econometrica* 47(1): 153-161.

Hox, J. J. 2010. *Multilevel Analysis: Techniques and Applications, 2nd Edition*. Routledge. p3

Hu, M. G., and J. F. Wang. 2011. A meteorological network optimization package using MSN theory. *Environmental Modelling & Software* 26: 546-548.

Hu, M. G., J. F. Wang, and Y. Zhao. 2013. A B-SHADE based best linear unbiased estimation tool for biased samples. *Environmental Modelling & Software* 48: 93-97.

Isaaks, E., and R. Srivastava. 1989. *Applied Geostatistics*. Oxford University Press.

Kolasa, J., C. D. Rollo. 1991. Introduction: The heterogeneity of heterogeneity: A glossary. 1-23 in: J. Kolasa & S. T. A. Pickett [eds.] *Ecological Heterogeneity*. Springer-Verlag, New York.

Kong, L. B., J. Y. Xin, W. Y. Zhang, and Y. S. Wang. 2016. The empirical correlations between $PM_{2.5}$, $PM_{10}$ and AOD in the Beijing metropolitan region and the PM2.5, PM10 distributions retrieved by MODIS. *Environmental Pollution* 216: 350-360.

Kulldorff, M. 1997. A spatial scan statistic. *Communications in Statistics: Theory and Methods*. 26: 1481-1496.

Lee, D., and R. Mitchell. 2013. Locally adaptive spatial smoothing using conditional auto-regressive models. *Applied Statistics* 62, Part4: 593-608.

Li, J. M., J. F. Wang, Z. P. Ren, D. Yang, Y. P. Wang, Y. Mu, X. H. Li, M. R. Li, Y. M. Guo, J. Zhu. 2020. Spatiotemporal trends in maternal mortality ratios in 2205 Chinese counties from 2010-2013 and ecological determinants: A Bayesian modelling analysis. *PLOS Medicine* 17(5): e1003114.

Lindley, D. V., M. R. Novick. 1981. The role of exchangeability in inference. *Annals of Statistics* 9: 45-58.

Liu, T. J., J. F. Wang, C. Xu, J. Q. Ma, C. D. Xu, and H. Y. Zhang. 2018. Sandwich mapping of rodent density in Jilin Province, China. *Journal of Geographical Sciences* 28(4): 445-458.

Lloyd, C. D. 2010. *Local Models for Spatial Analysis, 2nd Edition*. CRC.

Longley, P. A., M. F. Goodchild, D. J. Maguire, D. W. Rind. 2005. *Geographical Information Systems and Science, 2nd Edition.* John Wiley & Sons Ltd.

Matheron, G. 1963. Principles of Geostatistics. *Economic Geology* 58: 1246-1266.

MacEachren, A.M. 1982. Map complexity: Comparison and measurement. *The American Cartographer* 9: 1, 31-46.

Monmonier, M.S. 1974. Measures of pattern complexity for choroplethic maps. *The American Cartographer* 1: 2, 159-169.

O'Connell, P. E., R. J. Gurney, D. A. Jones, J. B. Miller, C. A. Nicholas, and M. R. Senior. 1979. A case study of rationalization of a rain guage network in SW England. *Water Resources Research* 15: 1813-22.

Openshaw S. 1984. *The Modifiable Areal Unit Problem*. CATMOG 38. Norwich: GeoAbstracts.

Osborne, P. E., G. M. Foody, S. Suárez-Seoane. 2007. Non-stationarity and local approaches to modelling the distributions of wildlife. *Diversity and Distributions* 13: 313–323.

Pearson, K. 1895. Notes on regression and inheritance in the case of two parents. *Proceedings of the Royal Society of London* 58: 240–242.

Rao, J. N. K. 2003. *Small Area Estimation*. New York: John Wiley.

Rao, J. N. K. 2014. *Small-Area Estimation*. Wiley StatsRef: Statistics Reference Online, 1-8.

Ripley, B. D. 1981. *Spatial Statistics*. Wiley.

Rodriguez-Iturbe, I., J. M. Mejia. 1974. The design of rainfall networks in time and space. *Water Resources Research* 10, 713–728.

Snijders, T.A.B., Bosker, R.J. 2011. *Multilevel Analysis: An Introduction to Basic and Advanced Multilevel Modeling, 2nd Edition*. Sage.

Sui, D. 2006. Tobler's First Law of Geography, In B. Warff (Ed.), *Encyclopedia of Human Geography*. Thousand Oaks, CA, Sage Publications, 2006: p.454.

Thompson, S. K. 2012. *Sampling, 3rd Edition*. Wiley.

Wang, J. F., Christakos, G., Hu, M. G. 2009. Modeling spatial means of surfaces with stratified non-homogeneity. *IEEE Transactions on Geoscience and Remote Sensing* 47(12): 4167-4174.

Wang, J. F., Haining, R. and Z. D. Cao. 2010a. Sample surveying to estimate the mean of a heterogeneous surface: reducing the error variance through zoning. *International Journal of Geographical Information Science* 24(4): 523-543.

Wang, J. F., X. H. Li, G. Christakos, Y. L. Liao, T. Zhang, X. Gu, X. Y. Zheng. 2010b. Geographical detectors-based health risk assessment and its application in the neural tube defects study of the Heshun Region, China. *International Journal of Geographical Information Science* 24(1): 107-127.

Wang, J. F., B. Y. Reis, M. G. Hu, G. Christakos, W. Z. Yang, Q. Sun, Z. J. Li, X. Z. Li, S. J. Lai, H. Y. Chen, D. C. Wang. 2011. Area disease estimation based on sentinel hospital records. *PLoS ONE* 6(8): e23428.

Wang, J. F., M. G. Hu, C. D. Xu, G. Christakos, Y. Zhao. 2013a. Estimation of citywide air pollution in Beijing. *PLoS ONE* 8(1): e53400.

Wang, J. F., R. Haining, T. J. Liu, L. F. Li, C. S. Jiang. 2013b. Sandwich estimation for multi-unit reporting on a stratified heterogeneous surface. *Environment and Planning A* 45(10): 2515-2534.

Wang, J. F., C. D. Xu, M. G. Hu, Q. X. Li, Z. W. Yan, P. Jones. 2018. Global land surface air temperature dynamics since 1880. *International Journal of Climatology* 38: e466-e474.

Wang, J. F., T. L. Zhang, B. J. Fu. 2016. A measure of spatial stratified heterogeneity. *Ecological Indicators* 67 (2016): 250-256.

Wang, J. X., M. G. Hu, B. B. Gao, H. M. Fan, J. F. Wang. 2019. A spatiotemporal interpolation method for the assessment of pollutant concentrations in the Yangtze River estuary and adjacent areas from 2004 to 2013. *Environmental Pollution*. DOI: 10.1016/j.envpol.2019.05.132.

Xu, C. D., J. F. Wang, M. G. Hu, Q. X. Li. 2013. Interpolation of missing temperature data at meteorological stations using P-BSHADE. *Journal of Climate* 26: 7452-7463.

Xu, C. D., J. F. Wang, Q. X. Li. 2018. A new method for temperatures spatial interpolation based on sparse historical stations. *Journal of Climate* 31: 1757-1770.

Xu, L., Liu, Q. Y., Stige, L. C., T. B. Ari, X. Y. Fang, K. S. Chan, S. C. Wang, N. C. Stenseth, Z. B. Zhang. 2011. Nonlinear effect of climate on plague during the third pandemic in China. *PNAS* 108(25): 10214-10219.

Yin, Q., J. F. Wang, Z. P. Ren, J. Li, Y. M. Guo. 2019. Mapping the increased minimum mortality temperatures in the context of global climate change. *Nature Communications* 10: 4640.

Zhang, D., Brecke, P., Lee, H.F., He, Y.Q., Zhang, J. 2007. Global climate change, war, and population decline in recent human history. *Proceedings of National Academy of Sciences of the United States of America* 104(49): 19214–9.